\documentstyle[12pt,twoside,fleqn,espcrc1,epsfig]{article}

\newcommand{\be}{\begin{equation}}
\newcommand{\ee}{\end{equation}}
\newcommand{\ba}{\begin{eqnarray}}
\newcommand{\ea}{\end{eqnarray}}
\newcommand{\ket}[1]{| {#1} \rangle}
\newcommand{\bra}[1]{\langle {#1}|}
\newcommand{\ave}[1]{\langle {#1} \rangle}
\newcommand{\mpi}{m_\pi}

\title{Theoretical Interpretations of Low-Mass Dileptons
  \thanks{work supported in part by BMBF, NSF-PHY94-12309 and  
   US-DOE under DE-FG02-88ER40388 .}}

\author{J. Wambach\address{Institut f\"ur Kernphysik,
Technische Universit\"at Darmstadt,\\
Schlo{\ss}gartenstr. 9, 
D-64289, Darmstadt, Germany}
and R. Rapp\address{Department of Physics and Astronomy,
State University of New York at\\
Stony Brook, NY 11794-3800, USA}
 \thanks{supported by the   
      A.-v-.Humboldt Foundation as a Feodor-Lynen fellow.}}

\begin{document}
\maketitle

\begin{abstract}

An overview is given of chiral symmetry restoration at finite
temperature and baryochemical potential. Within hadronic
models of the vector correlator its implications for
low-mass dilepton spectra in ultrarelativistic heavy-ion
collisions are discussed.

\end{abstract}

\section{INTRODUCTION}
\begin{figure}[htb]
\begin{center}
\epsfig{file=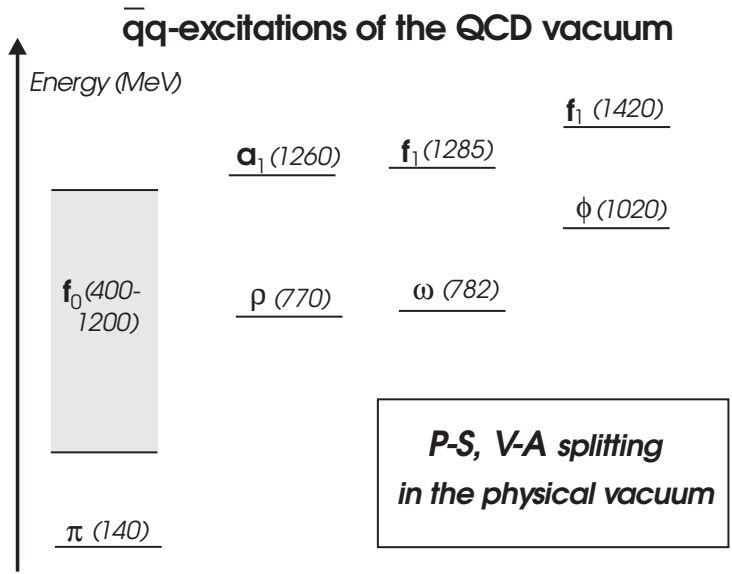,scale=0.4}
\caption{Low-lying spectrum of selected meson states.}
\label{fig:mspect}
\end{center}
\end{figure}

In the limit of vanishing quark masses QCD exhibits an
exact chiral symmetry, in which left- and right handed quarks
decouple giving rise to conserved vector- and axialvector
currents. In nature this limit is most relevant for the up- and down
quark flavors, where masses are small compared to typical
hadronic scales. To a somewhat lesser extent it also applies
to the three-flavor case $(N_f=3)$ which includes the strange
quark. Chiral symmetry is fundamental to our understanding
of light hadron masses where confinement seems to play a much
lesser role. 
In the physical vacuum chiral symmetry is spontaneously broken. For the 
light meson spectrum representing elementary quark-antiquark modes 
(Fig.~\ref{fig:mspect}), this manifests itself in two ways: 
\begin{itemize}
\item[--] the appearance of (nearly) massless Goldstone particles
          (pion, kaon, eta).
\item[--] the absence of parity dublets, i.e. the splitting of scalar- 
          and pseudoscalar, as well as vector- and axial vector mesons.
\end{itemize}
QCD sum rules relate meson masses to the quark condensate $\ave{\bar qq}_0$,
an order parameter of spontaneous chiral symmetry breaking. For the
$\rho$- and $a_1$-meson, for instance, one obtains in the large 
$N_c$-limit~\cite{Hats}
\be
m_\rho^2\approx \biggl [{448\over 27}\pi^3\alpha_s\ave{\bar q q}_0^2\biggr ]^{1/3}
\qquad 
m_{a_1}^2\approx \biggl [{2816\over 27}\pi^3\alpha_s\ave{\bar q q}_0^2\biggr ]^{1/3}
\, .
\label{sumr} 
\ee

\subsection{Chiral Symmetry Restoration}
It is expected that chiral symmetry is restored at finite temperature, $T$ and 
quark potential, $\mu_q$. The only reliable way to obtain the order of the 
transition, the critical temperature and the critical chemical potential 
is ab initio 
lattice simulations. While, at $\mu_q=0$, results are available with 
$T_c=150\pm 20$ MeV~\cite{Laer} there are difficulties for finite $\mu_q$. 
It has long been known that quenched QCD breaks down at non-zero $\mu_q$ in 
that the transition takes place at $\mu_q\approx\mpi/2$ which is unphysical. 
Within random matrix models this failure is now
understood as an inherent problem of the quenched approximation~\cite{Steph}. 
It is not clear at the moment whether an early onset
of the transition truly holds in full QCD or if
it reflects computational difficulties in the method employed for
the simulation~\cite{Ukawa}.
  
In spite of these difficulties with lattice QCD, model-independent results 
can be obtained at low temperatures and small baryochemical potential, $\mu$, 
from phenomenology. Considering the QCD-free energy 
\be
{\cal F}=-T\ln Z_{QCD}
\ee
where $Z_{QCD}$ is the grand-canonical QCD-partition function, the condensate
ratio can be expressed via the difference $\delta {\cal F}={\cal F}-{\cal F}_0$
as~\cite{GeLe}  
\be
{\ave{\bar qq}_{\mu,T}\over\ave{\bar qq}_0}=1-
{\partial\delta {\cal F}(\mu,T)\over \partial m_q} \, .
\label{ccrg}
\ee
Here $m_q$ denotes the quark masses, which act like the external field in a 
magnet.
For a dilute gas of hadrons the ratio can be expressed in the two-flavor case 
as~\cite{CEW}
\be
{\ave{\bar qq}_{\mu,T}\over\ave{\bar qq}_0}\approx 1-
\sum_h{\Sigma_h\rho^s_h(\mu,T)\over f_\pi^2 m_\pi^2}
\ee
where $\Sigma_h\propto \partial m_h/\partial m_q$ denotes the Sigma-Commutator
of hadron $h$ and $\rho^s_h$ its scalar density.
For low temperatures and small baryon density degenerate nucleons and  
thermally excited pions give the dominant contribution such that 
\be
\lim_{\mu\to 0, T\to 0} {\ave{\bar qq}_{\mu,T}\over\ave{\bar qq}_0}\sim
1-{1\over \mpi^2f_\pi^2}\biggl (\Sigma_\pi\rho^s_\pi(T)+
\Sigma_N\rho^s_N(\mu)\biggr ) \, .
\label{dilute}
\ee
Since $\Sigma_\pi=\mpi/2=69$ MeV and $\Sigma_N=45\pm 8$ MeV are known 
experimentally, the dilute gas limit is well determined.
The underlying physical picture is very simple. Whenever a hadron is
created in the vacuum the $\ave{\bar qq}$ condensate is changed locally
since the condensate inside a hadron is different from that in the vacuum.

At $\mu=0$, one can go a step further and derive a rigorous low-temperature
expansion by means of chiral perturbation theory~\cite{GaLe,GeLe}. 
Taking the chiral limit $\mpi\to 0$ one obtains 
\be
{\ave{\bar qq}_{T}\over\ave{\bar qq}_0}=1-x-
{1\over 6}x^2-{16\over 9}x^3 \ln{T\over\Lambda}+...\quad ;
\qquad x=\biggl ({T\over \sqrt{8}f_\pi}\biggr )^2
\ee
where the temperature scale is set by $\sqrt{8}f_\pi\sim 250$ MeV. 
While the $T^2$-and $T^4$-terms are model independent,
model dependence enters at order $T^6$ through the regularization scale 
$\Lambda\sim 470$ MeV. The range of validity is 
restricted to $T< 120$ MeV mostly because at this point 
heavier mesons start to enter. 

For finite $\mu$ the dilute gas expression (\ref{dilute}) predicts a
decrease of the condensate ratio which is linear in the vector 
density (for heavy particles the scalar density becomes equal to the
vector density). At nuclear saturation, 
$\rho_0=0.16/fm^3$, this yields a $\sim 35 \%$ drop and naive
extrapolation would indicate chiral restoration at $\sim 3\rho_0$. This
clearly cannot be trusted, since the equation of state of nuclear
matter greatly differs from that of a free Fermi gas at such high
densities~\cite{Weise}.

\subsection{How to Detect Symmetry Restoration?}
The principal tool for observing  QCD phase transitions are ultrarelativistic
heavy-ion collisions (URHIC's). In the study of chiral restoration the quark 
condensate $\ave{\bar qq}_{\mu,T}$ is not directly measurable, however, and 
one has to look for other observables.
It follows from chiral symmetry alone that, at the phase boundary,
the scalar and pseudo-scalar correlators as well as the vector- and axialvector 
correlators must become identical. In principle, the observation of 
parity mixing can 
thus serve as a unique signal for chiral restoration. 
The study of phase transitions via correlators is known in condensed
matter physics, as 'soft-mode' spectroscopy~\cite{Hats}. Here structural phase transitions
in crystals are analyzed by measuring the frequency of low-lying lattice 
phonons ('soft modes'). A good example is $SrTiO_3$~\cite{Gebh}.
Above a critical a temperature $T_c=106$ K this crystal has cubical symmetry 
with a three-fold degenerate '$R_{25}$-mode'. Upon cooling this mode
becomes soft with vanishing frequency at $T_c$.  At this point the crystal 
undergoes a structural transition to a phase of tetragonal symmetry. 
As a consequence the three-fold degenerate soft mode splits into two modes
one of which is two-fold degenerate. Upon further cooling the
frequencies of these modes as well as their splitting increase. Obviously
the low-lying phonon spectrum contains the information of the 
underlying symmetry change during a phase transition.

The direct QCD-analog is the behavior of the scalar- 
and pseudoscalar correlators.
As manifest in the linear $\sigma$-model they are
the carriers of chiral symmetry. As $T_c$ is approached from above, 
the scalar ($\sigma$)- and pseudoscalar ($\pi$) modes become soft 
and split below
$T_c$ with the pion remaining massless in the chiral limit. The measurement
in heavy-ion collisions
of the scalar- and pseudoscalar correlators is difficult, however, since
pions suffer strong final-state interactions. A better
probe is the vector correlator since it couples 
to photons or dileptons which only undergo  electromagnetic 
final-state interaction. 
Since the vector correlator in the 
vacuum is saturated by narrow 'resonances', the $\rho,\omega$ and 
$\phi$-mesons, the objective is then to study the spectral changes of 
vector mesons as a function
of $\mu$ and $T$. There are, in principle, two possibilities. As the scalar
and pseudoscalar modes, also the vector modes could become 'soft' at $T_c$,
giving rise to 'dropping masses'. This is the hypothesis 
of 'Brown-Rho scaling'~\cite{BR91} and would be a natural 
consequence of a direct relationship between 
the masses and the chiral condensate, as found in the vacuum 
(see eq.~(\ref{sumr})). Brown-Rho scaling
also explains qualitatively the rapid increase in entropy density across
the phase boundary, seen in lattice QCD~\cite{Karsch}.  The second possibility 
is that the vector mesons remain massive at $T_c$, becoming degenerate 
with their axial partners. Support for this possibility comes from the fact 
that the masses of the $\rho$- and $a_1$-meson change as 
$T^4$~\cite{Elets} which is difficult to 
reconcile from a direct link between masses and the chiral condensate in the
medium. Recall that the condensate ratio, in the chiral limit, decreases as
$T^2$, for small $T$.

\section{HADRONIC DESCRIPTIONS OF THE VECTOR CORRELATOR}
\subsection{General properties}
The vector correlator in the vacuum is defined as the current-current
correlation function 
\be
  \Pi_{\mu\nu}^0(q)=i\int d^4x e^{iq\cdot x}
  \bra{0}{\cal T}j_{\mu}(x) j_{\nu}(0)\ket{0}  ,
 \label{2.1}
\ee 
where ${\cal T}$ denotes the time-ordered product and $j_{\mu}$ is the
electromagnetic current. For three flavors the current can be decomposed in
terms of the quantum numbers of the relevant vector mesons as 
\be
  \label{2.1b}
  j_{\mu}= j_\mu^{(\rho)}+j_\mu^{(\omega)}+j_\mu^{(\phi)}\, .
\ee
The corresponding quark content is
\ba
\label{2.2a}
  j^{(\rho)}_\mu&=&\frac{1}{2}(\bar{u}\gamma_{\mu}u-\bar{d}\gamma_{\mu}d),\\
\label{2.2b}
  j^{(\omega)}_\mu&=&\frac{1}{6}(\bar{u}\gamma_{\mu}u+\bar{d}\gamma_{\mu}d),\\
  j^{(\phi)}_\mu&=&-\frac{1}{3}(\bar{s}\gamma_{\mu}s).
\label{2.2c}
\ea
Current conservation implies a transverse tensor structure
\be
  \Pi_{\mu\nu}(q)=\left(g_{\mu\nu}-\frac{q_{\mu}q_{\nu}}{q^2}\right)\Pi (q^2),
\qquad 
  \Pi (q^2)=\frac{1}{3} g^{\mu\nu}\Pi_{\mu\nu}(q).  
\label{2.3}
\end{equation}
The imaginary part of $\Pi (q^2)$ is proportional to the cross section for 
$e^+ e^- \to$ hadrons 
\be
 R(s)=\frac{\sigma(e^+ e^-\to \rm hadrons )}{\sigma (e^+
  e^-\to \mu^+ \mu^-)}=-\frac{12 \pi}{s} {\rm Im} \Pi(s),
\label{2.6}
\ee
where $\sqrt{s}$ is the total c.m. energy of the lepton pair. The
vector mesons appear as resonances in the different hadronic channels
carrying the respective flavor quantum numbers and the vector dominance model 
(VDM) quantitatively describes the resonant contribution to
$R(s)$~\cite{Kling}.

In a hadronic medium the vector correlator looses its simple vacuum
Lorentz structure due to a preferred frame.
In the grand canonical ensemble one has
\be
\Pi_{\mu\nu}^{(\mu,T)}(q_0,\vec q)=i\int d^4x \: e^{iq\cdot x}
{\rm Tr}\biggl (e^{(-\hat H-\mu \hat N)/T}j_{\mu}(x) j_{\nu}(0)\biggr )/Z
 \label{2.8}
\ee
with separate dependence on energy and three-momentum. In addition, the 
longitudinal and transverse components of $\Pi_{L,T}^{(\mu,T)}$ 
now become distinct. The same holds for the propagators
\be
D^{(\mu,T)}_{L,T}=(q^2-(m_V^0)^2-\Pi^{(\mu,T)}_{L,T})^{-1}\, .
\ee

\subsection{Theoretical approaches}
At SpS energies the phase space in the final state is dominated by mesons 
(mostly pions) with a meson/baryon ratio of 5-7~\cite{Stach}. In calculating
modifications of vector meson properties through hadronic interactions, an
obvious starting point is therefore a pure pion gas. For the $\rho$ meson
which in the VDM couples dominantly to two-pion states this implies a modification
of the pion loop through the heat bath. The resulting thermal
broadening has been calculated e.g. for on-shell $\rho$ mesons~\cite{Hagl} 
or within the hidden gauge approach~\cite{SoKo}
and found to be rather small. Accounting for off-shell propagation in 
e.g. resonant $\pi\rho$-scattering through the $a_1(1260)$ meson 
gives rise to a  $\pi a_1$-loop in the $\rho$-meson self energy. Especially 
for energies above 1 GeV this contribution might be substantial, depending,
however, crucially on the choice of effective Lagrangian~\cite{GaGa}. 

\begin{figure}[htb]
\begin{minipage}[t]{0mm}
{\makebox{\epsfig{file=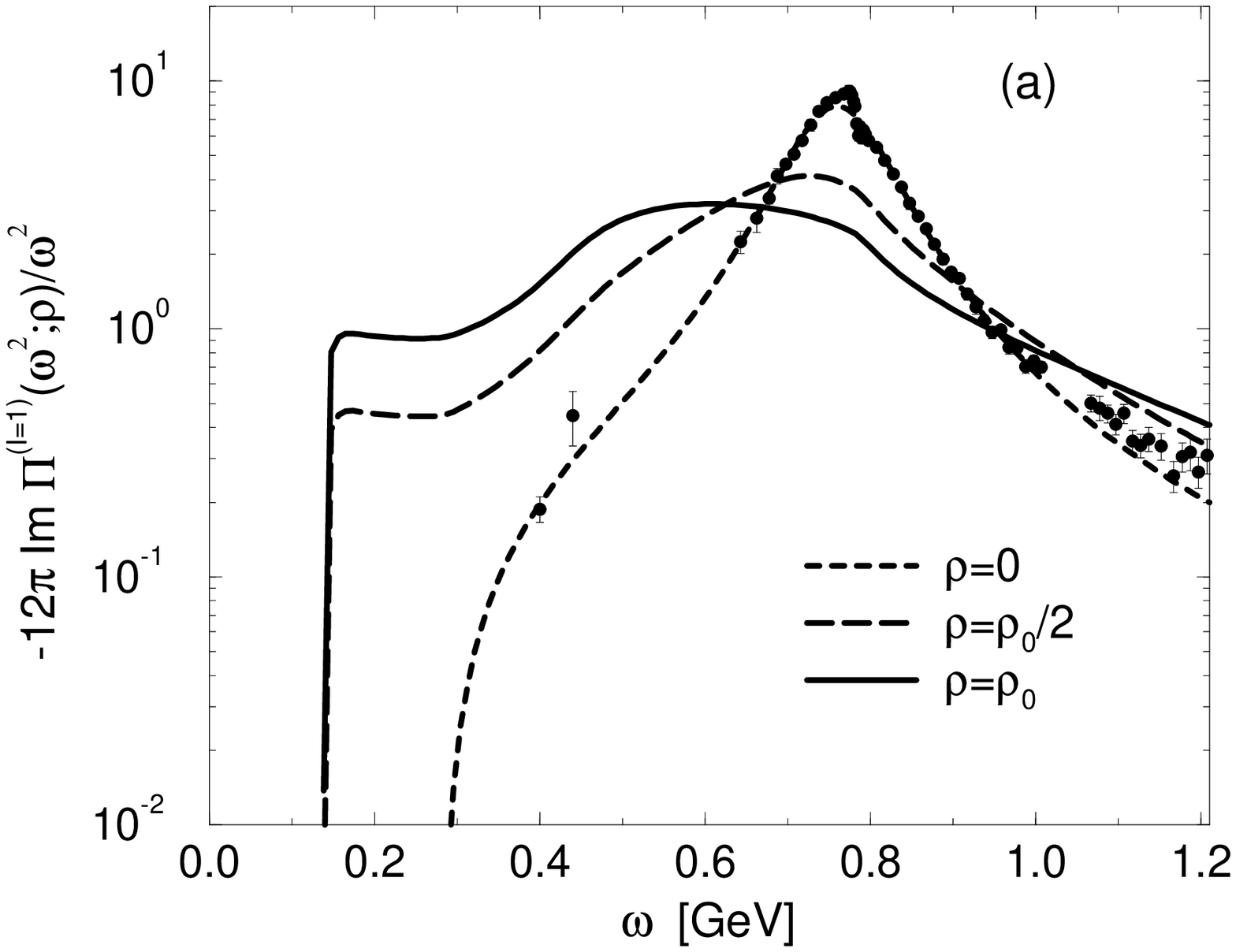,width=55mm}}}
\end{minipage}
\hspace{5cm}
\begin{minipage}[t]{0mm}
{\makebox{\epsfig{file=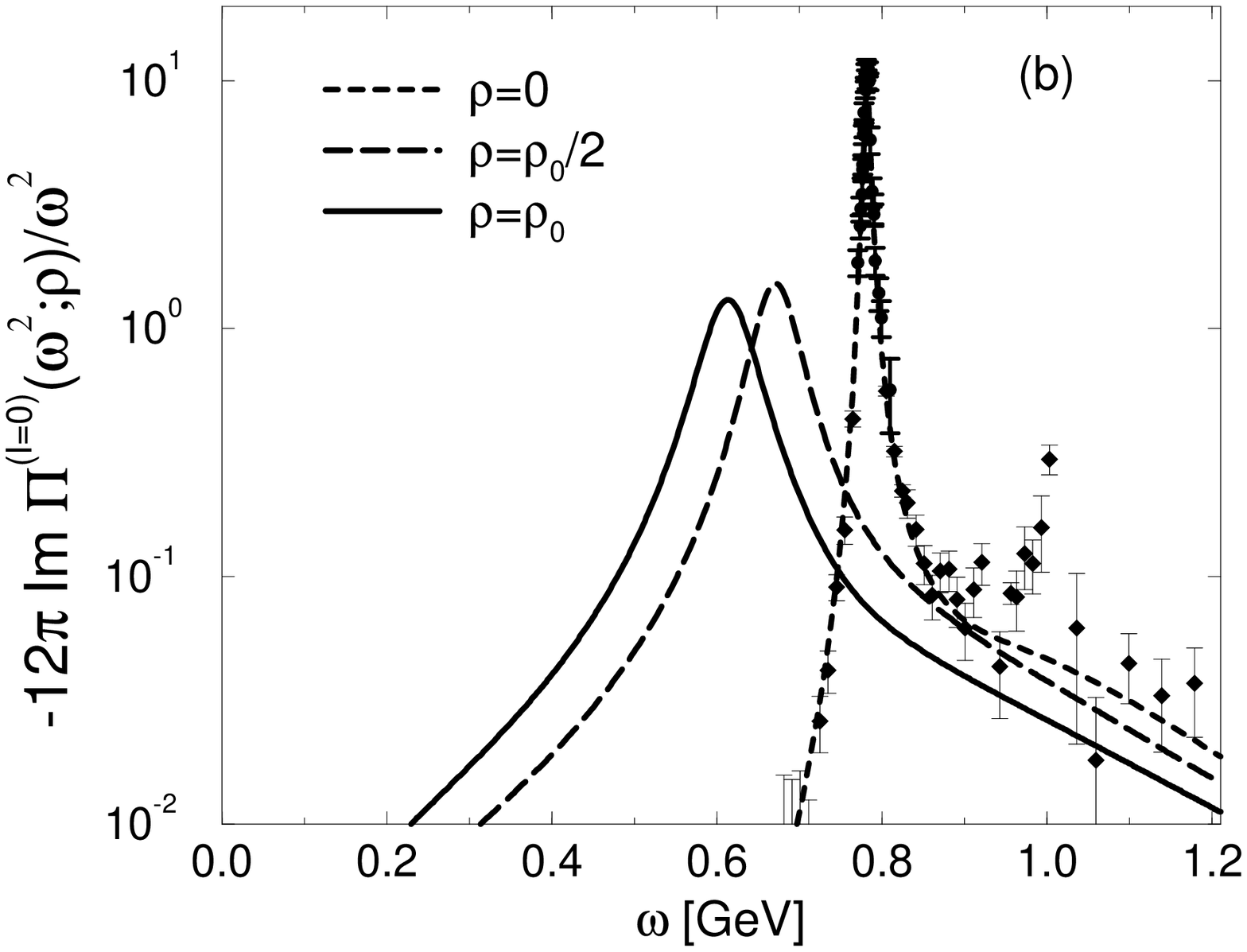,width=55mm}}}
\end{minipage}
\hspace{5cm}
\begin{minipage}[t]{0mm}
{\makebox{\epsfig{file=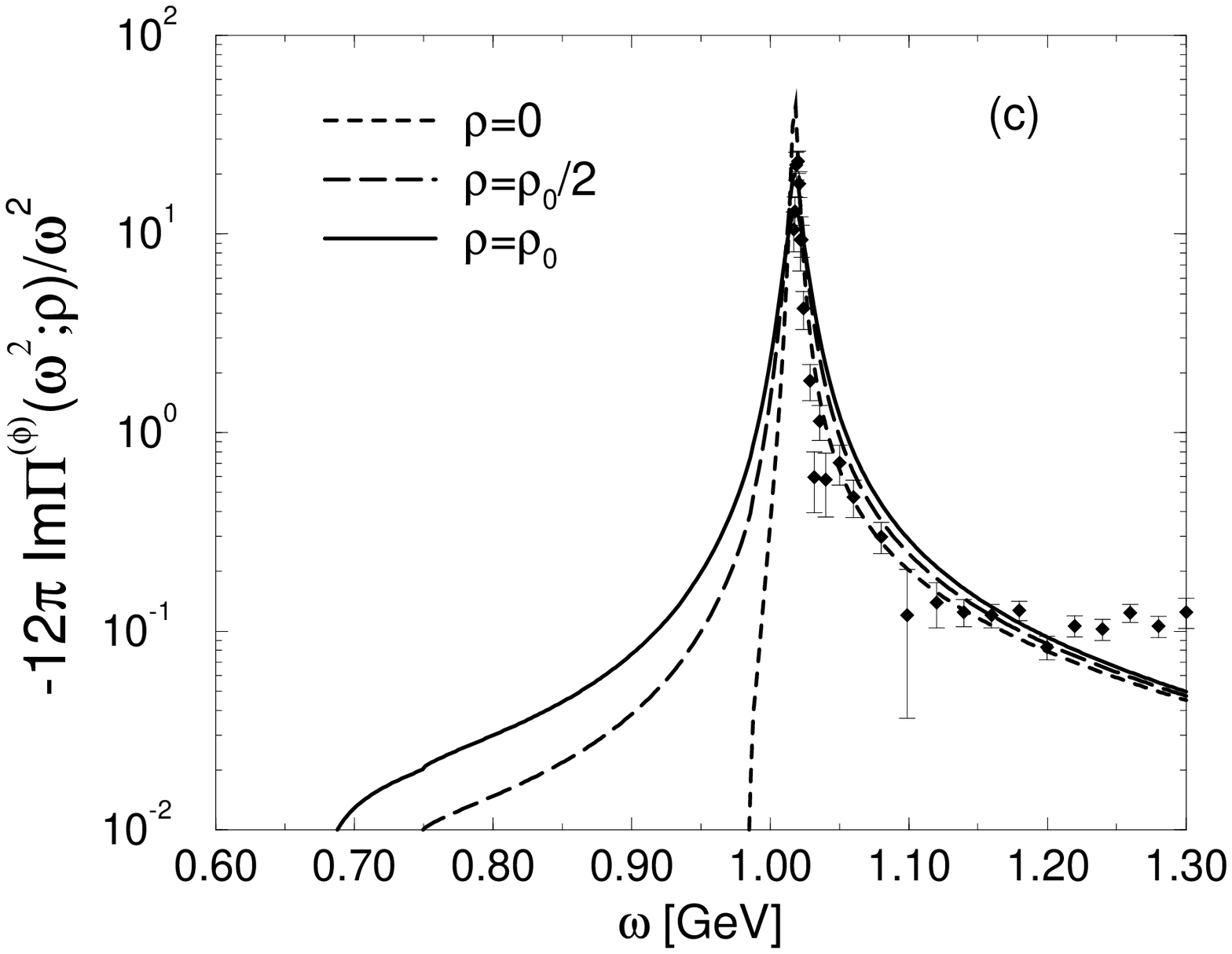,width=55mm}}}
\end{minipage}
\caption{Modifications of the vector-meson spectrum~[13].
Panels (a), (b) and (c) depict the $\rho$-, $\omega$- and $\phi$ meson,
respectively, The short-dashed lines denote the vacuum values, while
the dashed (full) lines indicate the results at $\rho_0/2$ ($\rho_0$).}
\label{fig:specm}
\end{figure}

In spite of the scarcity of baryons in the hot hadron gas, they have a
significant impact on the spectral properties of vector mesons, largely
because of strong meson-baryon coupling. Several approaches have been
put forward to determine these effects reliably. One approach starts from
a low-density expansion of the vector correlator~\cite{Kling,Steele}. 
To leading order in density the vector correlator takes the form
\be
\Pi_{\mu\nu}^{(\mu,T)}(x)=\Pi_{\mu\nu}^0+(x)+\bra{\pi}
{\cal T}j_{\mu}(x) j_{\nu}(0)\ket{\pi}\rho_\pi(T)+  
\bra{N}{\cal T}j_{\mu}(x) j_{\nu}(0)\ket{N}\rho_N(\mu)+..
\label{low}
\ee
and involves the pion- and nucleon Compton amplitudes. In the work
of Steele et al.~\cite{Steele} these are obtained from data 
in a 'chiral reduction formalism' and automatically 
contain mixing of the vacuum vector- and axialvector correlators 
in the pion sector, as implied by chiral symmetry. The nucleon Compton
amplitude is constrained from $\gamma N$ photoabsorption data and
the nucleon polarizabilities. 
In ref.~\cite{Kling} the $T=0$ limit of the low-density expansion (\ref{low})
is considered.  Then only the nucleon contributes and
its Compton tensor is evaluated in the VDM combined with chiral $SU(3)$
dynamics, based on an effective meson-baryon Lagrangian.
In the heavy-baryon formalism, the vector meson-baryon amplitude is
calculated for $\rho,\omega$- and $\phi$ meson scattering and significant
modifications of the spectral properties are found (Fig.~\ref{fig:specm}). 
Especially
the $\rho$ meson suffers substantial broadening, as density increases.
The position of the 'pole mass' is not affected, however. The in-medium
$\omega$ meson also broadens and exhibits a downward shift of the 'pole
mass'. The $\phi$ meson spectral function
is mostly affected in the sub-threshold region.

A second approach, which has been applied so far only for the $\rho$ meson, 
starts from the well-known observation that pion propagation 
in the nucleus is strongly modified. A wealth of elastic $\pi$-nucleus
scattering data has provided detailed understanding of the relevant
physical mechanisms~\cite{ErWe}. The dominant contributions originate
from non-resonant and resonant $\pi N$ scattering. Thus the pion
self energy acquires large contributions from $N$-hole and $\Delta$-hole
loops, giving rise to a momentum-softening of the pion dispersion
relation. In the VDM it is therefore natural to account
for this effect by replacing the vacuum two-pion loop with in-medium
pions~\cite{HeFN,AKLQ,ChSc}. Gauge invariance is ensured by including 
appropriate vertex corrections. To lowest order in nucleon density 
this approach represents a pion cloud model for the nucleon Compton amplitude
which coincides with that of ref.~\cite{Steele} and ref.~\cite{Kling}
(in the latter case there are additional box diagrams). In addition
to leading-order contributions in $\rho_N$ there naturally emerge higher
orders in density, most notably $\rho_N^2$ terms.
These correspond to two-nucleon processes of meson-exchange character
and $NN$- and $N\Delta$-bremsstrahlung contributions.
Besides the medium modifications caused by dressing the intermediate 
2-pion states, direct interactions of the $\rho$ meson with 
surrounding nucleons in the gas have to be considered. While elastic
scattering, $\rho N \to N \to \rho N$, is kinematically strongly disfavored
such a kinematic suppression will be much less pronounced with increasing 
energy of the resonance in the intermediate state. 
\begin{table}[b]
\label{tab1}
\caption{Baryon resonances with appreciable decay width into $\rho N$.}
\begin{center}
\begin{tabular}{ccc}
\hline
 $B^*$ & $I(J^p)$ & $\Gamma_{\rho N}$ [MeV] \\
\hline
$N$(1520)      & 1/2(3/2$^-$) &  24   \\
$\Delta$(1620) & 3/2(1/2$^-$) &  22.5 \\
$\Delta$(1700) & 3/2(3/2$^-$) &  45   \\
$N$(1720)      & 1/2(3/2$^+$) & 105   \\
$\Delta$(1905) & 3/2(3/2$^+$) & 210   \\
\hline
\end{tabular}
\end{center}
\end{table}  
Indeed, there are at 
least two well-established resonances in the particle data table~\cite{PDG} 
which strongly couple to the $\rho N$ decay channel, namely the 
$N$(1720) and the $\Delta$(1905). 
This led to the suggestion~\cite{FrPi} to consider $\rho$-like particle-hole 
excitations of the type $\rho N(1720)N^{-1}$ and 
$\rho\Delta(1905)N^{-1}$. In a more complete description other  
resonances with appreciable $\rho N$ widths have been 
included~\cite{RCW,PPLLM,RUBW}.
These are listed in Tab.~1.
The most simple version of the VDM overestimates the 
$B^*\to N\gamma$ branching fractions when using the hadronic coupling 
constants deduced from the $B^*\to N\rho$ partial widths. This is  
corrected for by employing an improved version of the VDM~\cite{KLZ},
which allows to adjust the $B^*N\gamma$ coupling independently.

\begin{figure}[htb]
\begin{minipage}[t]{0mm}
{\makebox{\epsfig{file=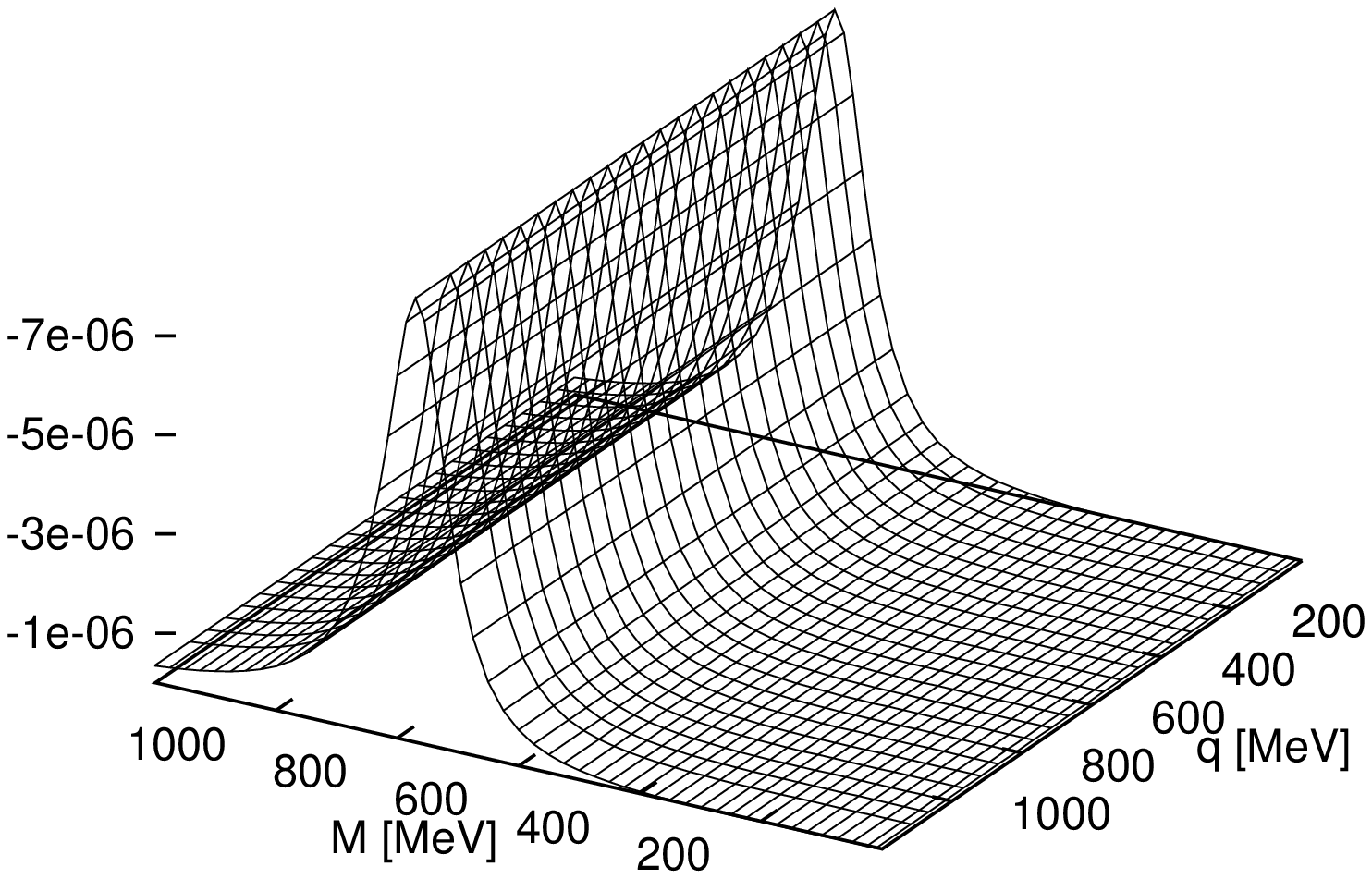,width=45mm,angle=-90}}}
\end{minipage}
\hspace{5cm}
\begin{minipage}[t]{0mm}
{\makebox{\epsfig{file=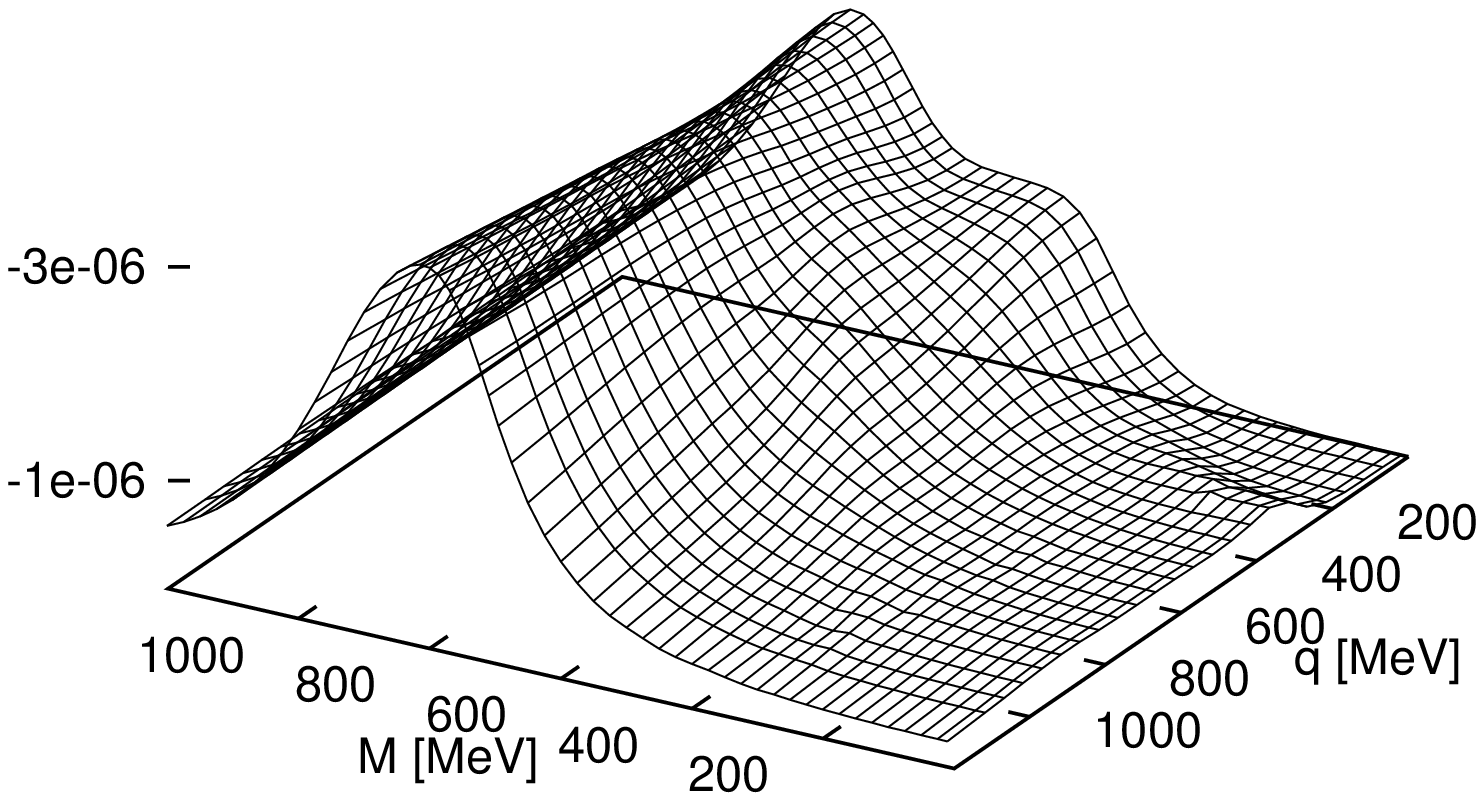,width=45mm,angle=-90}}}
\end{minipage}
\hspace{5cm}
\begin{minipage}[t]{0mm}
{\makebox{\epsfig{file=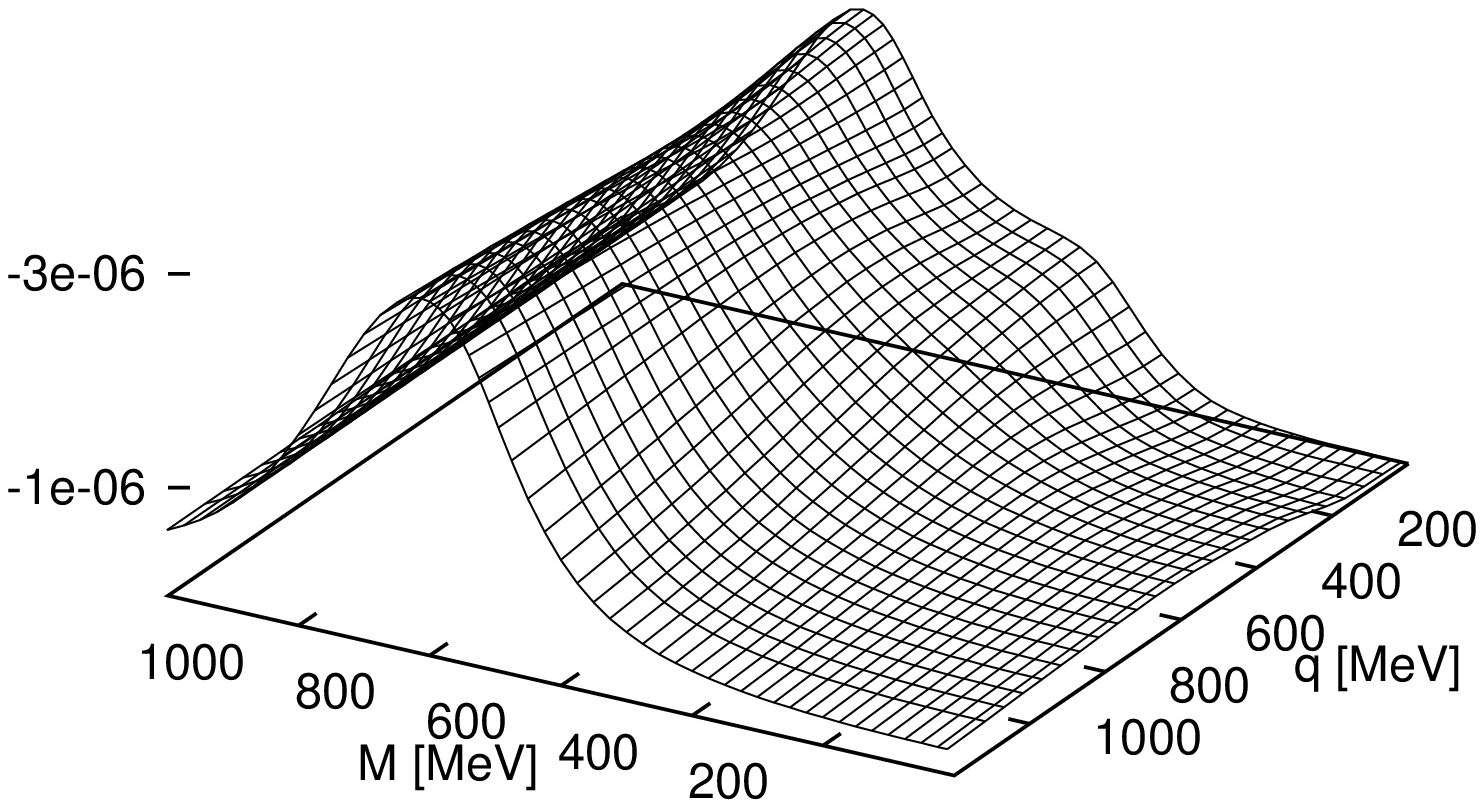,width=45mm,angle=-90}}}
\end{minipage}
\caption{The $\rho$-meson spectral function~[25,27] in the vacuum 
(left panel), at $\rho_N=\rho_0,T=0$ (middle panel) and at 
$\rho_N+\rho_{\Delta}=1.4\rho_0,T=166$~MeV (right panel); notice that the
vertical scale in the middle and right panel is reduced by a factor of 2.}
\label{fig:matten}
\end{figure}

Combining the effects of pionic modifications and resonant $\rho N$
scattering, the resulting $\rho$-meson spectral functions 
${\rm Im} D^\rho=1/3(D_L^\rho+2D_T^\rho)$ are displayed
in Fig.~\ref{fig:matten}.

One observes significant broadening especially at small $q$ with
a low-mass shoulder which originates from intermediate $\Delta N^{-1}\pi$
states and resonant $N(1520)N^{-1}$ excitations.
As $q$ increases the shoulder moves towards the $M=0$ line, which
can be understood from simple kinematics. As seen from
the right panel, temperature has little influence on the energy-momentum
dependence and only provides an additional overall broadening.

\subsection{Constraints from photoabsorption}
An obvious constraint for models of $\rho$-meson propagation
is photoabsorption in nuclei, for which a wealth of  
data exist over a wide range of energies. Real photons correspond to the 
$M=0$ line in the middle part of Fig.~\ref{fig:matten}. Within the
model discussed above, the total photoabsorption cross section per
nucleon can be calculated straightforwardly.
Taking the low-density limit, $\rho_N\to 0$, only terms
linear in density contribute, representing the absorption process
on a single nucleon. Two contributions have to be distinguished:
(i) coupling of the photon to the virtual pion cloud of the nucleon,
incorporated in the VDM via pion dressing; this contribution describes 
the background contribution the absorption process, as mentioned above;
(ii) resonant contributions via the intermediate $B^*$ resonances listed
in Tab.~1.

\begin{figure}[htb]
\begin{minipage}[t]{0mm}
{\makebox{\epsfig{file=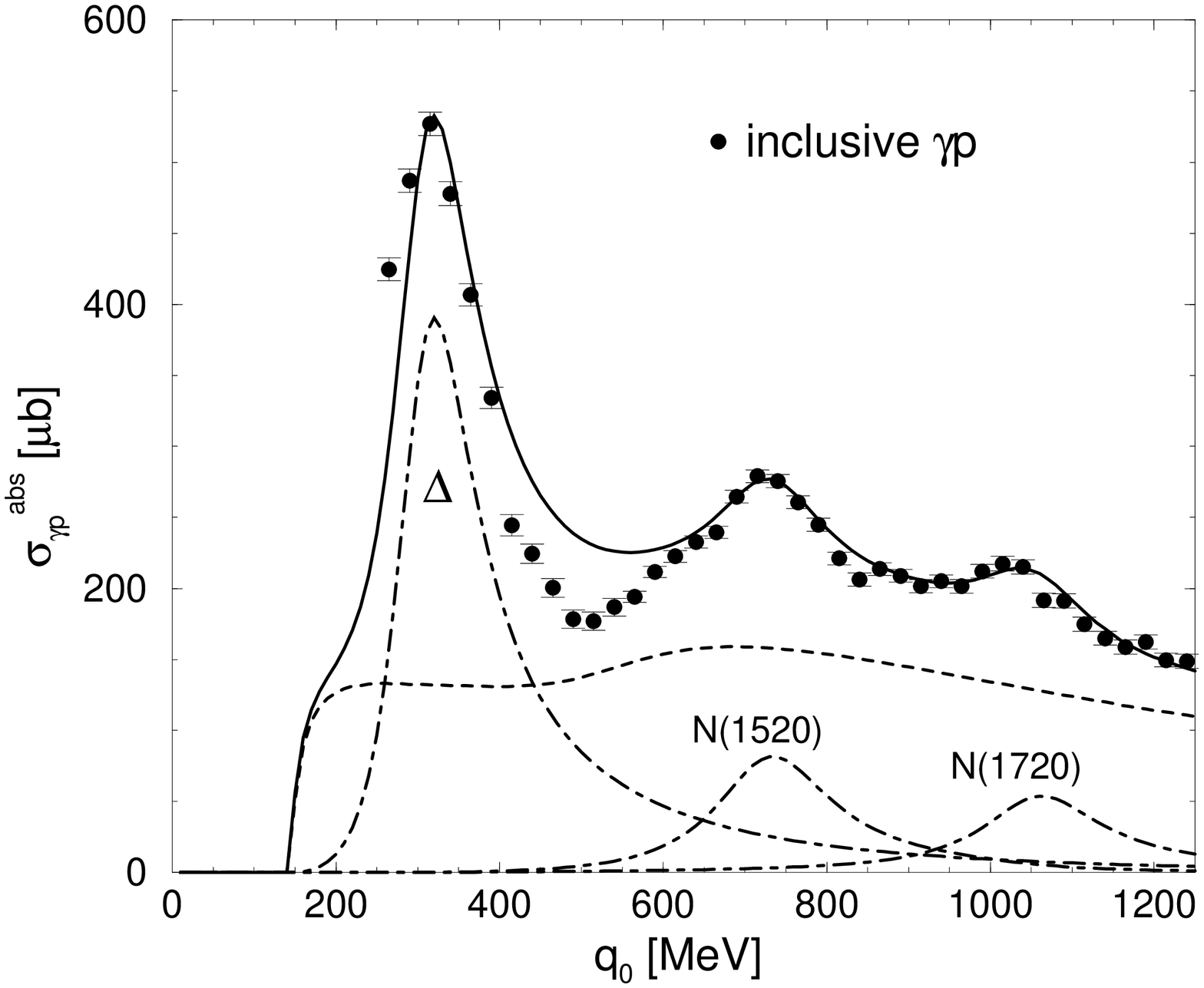,width=65mm, angle=-90}}}
\end{minipage}
\hspace{7cm}
\begin{minipage}[t]{0mm}
{\makebox{\epsfig{file=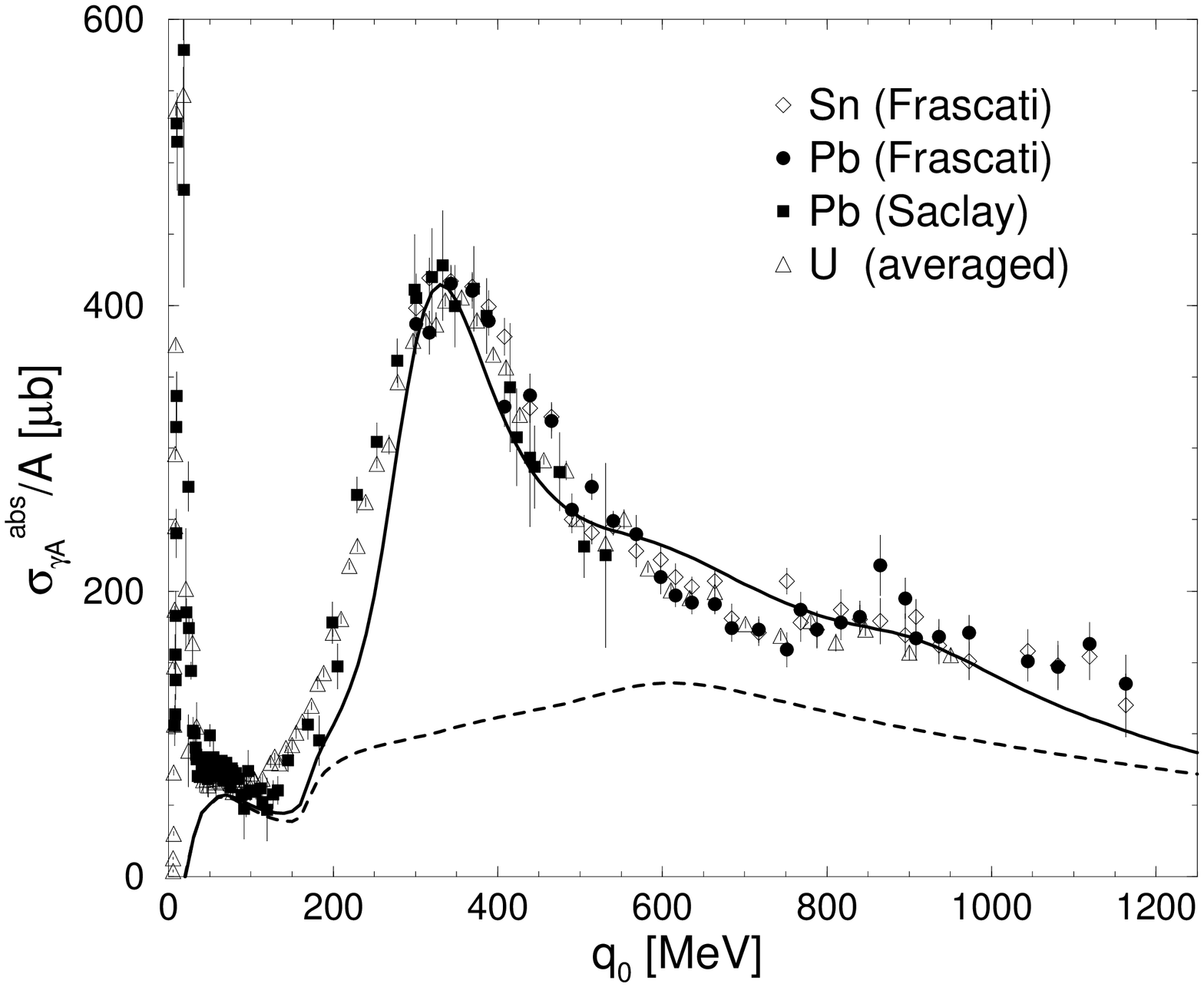,width=65mm, angle=-90}}}
\end{minipage}
\caption{The photo absorption spectrum on the proton (left panel)
and on nuclei (right panel) as obtained in ref.~[27]. The
dashed lines indicate the non-resonant background contributions.
The data were taken from refs.~[29-33].}
\label{fig:photo}
\end{figure}

Adjusting the model parameters to optimally reproduce the $\gamma p$
data~\cite{RUBW} yields results displayed in the left panel of 
Fig.~\ref{fig:photo}. Photoabsorption on nuclei can be reproduced with
similar quality.
It is noteworthy that, in the nucleus, strength below $m_\pi$ is obtained.
which originates from two-nucleon processes via meson-exchange currents and is
nothing but the 'quasi-deuteron tail' of the giant dipole resonance.

\subsection{Comparison with Dilepton data}

Constraining hadronic models via photoabsorption~\cite{Steele,RUBW} gives
confidence in extrapolations to the time-like region of dilepton production.
For the $\rho$ meson the dilepton rate is obtained as
\be
{dN_{\pi^+\pi^-\to l^+l^-}\over d^4xd^4q} =
-\frac{\alpha^2 (m_\rho^0)^4}{3\pi^3 g^2} \
\frac{f^\rho(q_0;T)}{M^2} \ g^{\mu\nu} \ {\rm Im}D^\rho_{\mu\nu}
(q_0,q;\mu,T)\, .
\label{rate}
\end{equation}
In the of model~\cite{RCW} most of the important processes
discussed above have been included such that the $\rho$-meson propagator
becomes
\be
D^\rho_{\mu\nu}=D^{\rho\pi\pi}_{\mu\nu}+D^{\rho NB^*}_{\mu\nu}+
D^{\rho\pi a_1}_{\mu\nu}+D^{\rho K K_1}_{\mu\nu} 
\ee
and contains aside from the baryonic contributions also the most
important parts from the pion/kaon gas.

To compare the theoretical rates with data, the space-time history
of the heavy-ion collision has to be specified. There are
several possibilities for modeling the collision. The most naive
approach is a simple 'fireball' model~\cite{CRW,RCW}, in which initial 
conditions in temperature and hadron abundances 
are obtained from transport model calculations. 
Assuming local thermal
equilibrium as well as chemical equilibrium, the space-time history
is then determined by a simple cooling curve, $T(t)$, from some
initial time, $t_i$, up to the 'freeze-out time', $t_f$.
\begin{figure}[!t]
\begin{center}
{\makebox{\epsfig{file=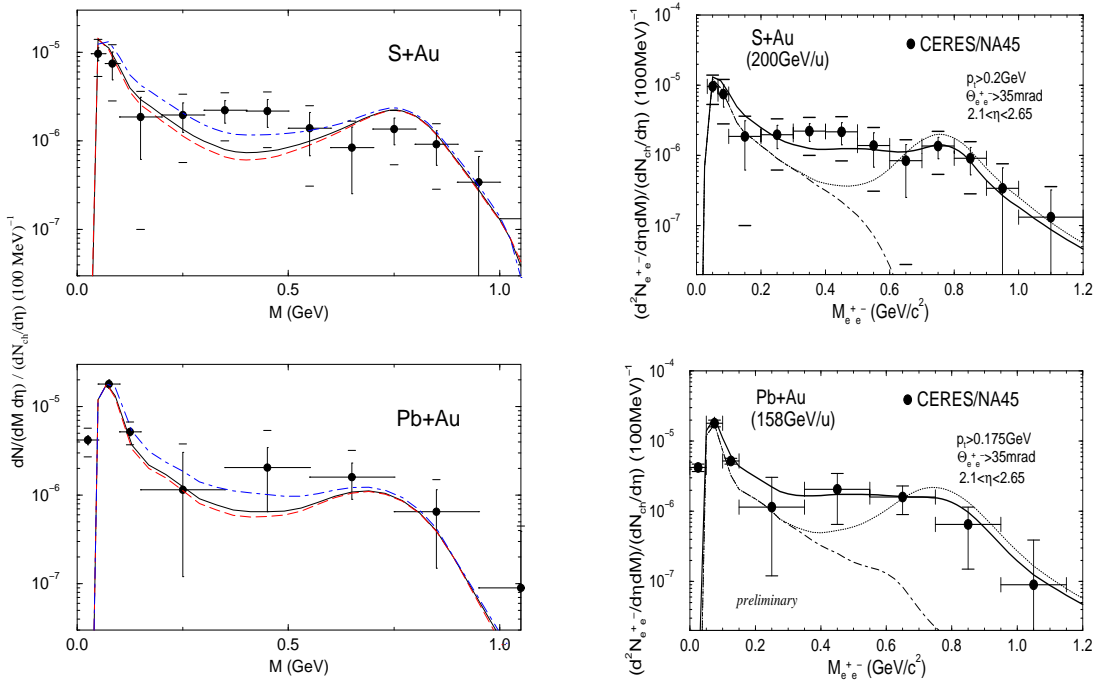,width=150mm,height=120mm}}}
\caption{CERES Dilepton Spectra~[36] employing the expanding fireball 
model of ref.~[25] with in-medium dilepton rates from ref.~[18]
(full lines in left panels) and ref.~[25] (full lines in right panels).}
\label{fig:drat1}
\end{center}
\end{figure}
 
For SpS
energies such cooling curves are available from transport theory~\cite{LKB} 
and can be easily parameterized. The time evolution of the hadron abundances 
is determined by chemical equilibrium and agrees well with 
transport model results. The observed spectrum is then obtained by 
integrating the 'local rate' (\ref{rate}) in time, accounting for the
detector acceptance in addition.
Results from~\cite{RCW,Steele} are displayed in Fig.~\ref{fig:drat1}.
While the full results in the right panel~\cite{RCW} give a reasonable account
of the observed spectra, especially in the region below the $\rho,\omega$
mass, the results of ref.~\cite{Steele} fail to describe this region.
The reason for this discrepancy between the two calculations is
currently not completely understood.
\begin{figure}[htb]
\begin{minipage}[t]{0mm}
{\makebox{\epsfig{file=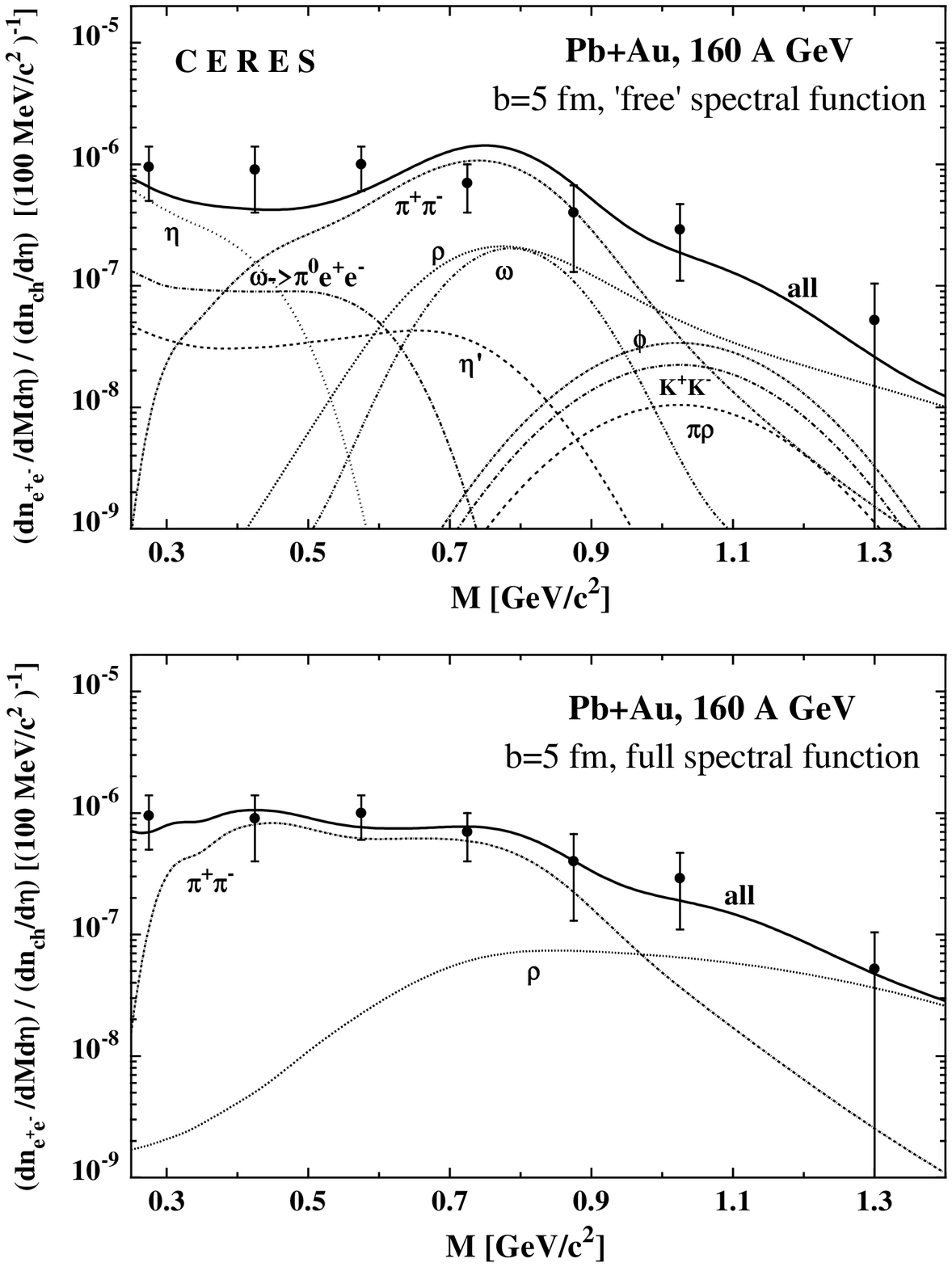,width=85mm}}}
\end{minipage}
\hspace{7cm}
\begin{minipage}[t]{0mm}
{\makebox{\epsfig{file=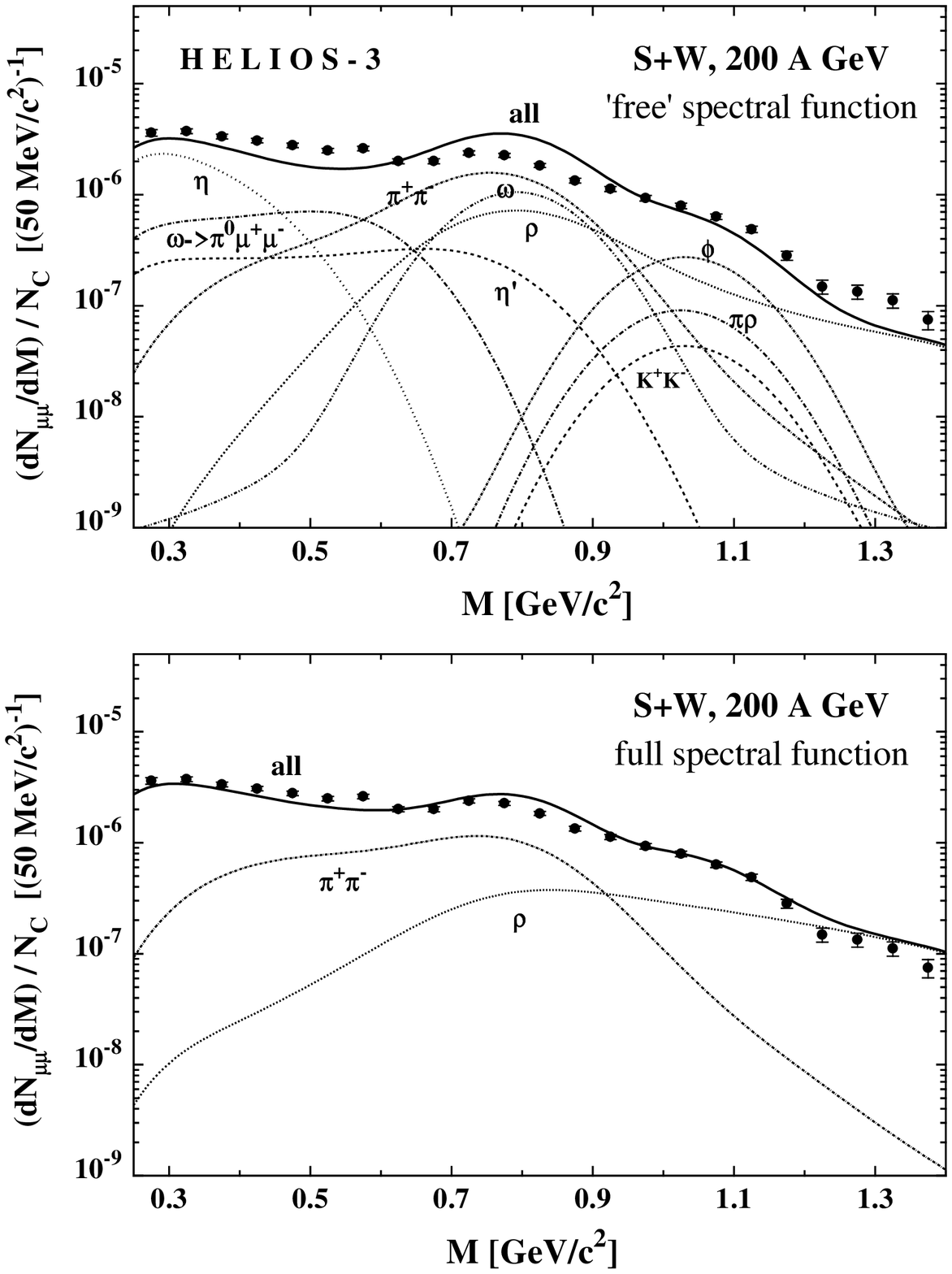,width=85mm}}}
\end{minipage}
\caption{SpS Dilepton Spectra~[36,38] employing the HSD transport model~[37] 
using free (upper panel) and  in-medium~[25,27]  (lower panel) $\rho$-meson 
spectral functions.}
\label{fig:drat2}
\end{figure}

The fireball model is rather crude and does not incorporate detailed 
flow dynamics. A more satisfactory description is provided
by hydrodynamical simulations. In this case the 'local rates' (\ref{rate})
refer to a given fluid cell and can be directly implemented. So far, however,
no results are available.

A third way is to supply transport calculations with rates 
obtained from in-medium vector-meson propagation. For the $\rho$ meson
this has been implemented in simulations using the $HSD$ model for
the transport~\cite{CBRW}. Results have been obtained for both the CERES
$e^+e^-$-measurement in Pb+Au collisions (left panel of Fig.~\ref{fig:drat2})
as well as the HELIOS-3 $\mu^+\mu^-$-measurement of S+W collisions
(right panel of Fig.~\ref{fig:drat2}). In both cases, the in-medium rates give 
an improvement in the comparison between theory and experiment and the
overall description is quite satisfactory.

\section{SUMMARY}

Recent theoretical efforts in understanding the nature of chiral 
symmetry restoration at
finite temperature and baryochemical potential and their implications for
low-mass dilepton production in URHIC's have been discussed. While the
onset of chiral symmetry restoration is well understood in the dilute gas
limit, the approach towards the phase boundary within hadronic models remains
a theoretical challenge. Here symmetry-conserving nonperturbative methods
are called for. A unique signal for chiral symmetry restoration in URHIC's
would be the observation of degenerate vector- and axialvector correlators.
Such a mixing strictly follows from chiral symmetry but is
difficult to detect. Only the vector correlator is accessible
via electromagnetic probes.

The application of hadronic models to assess in-medium properties of vector 
mesons and their consequences for dilepton spectra have reached 
a degree of maturity such that realistic comparisons with data 
become meaningful. While some approaches
inherently involve constraints from chiral symmetry~\cite{Kling,Steele}, 
others lack an obvious connection to chiral symmetry emphasizing, however,
input from hadronic phenomenology~\cite{CRW,RCW,RUBW}.  
When supplemented by models for the space-time history of the collision, 
quantitative agreement between theory and experiment seems to  
emerge~\cite{RCW,CBRW}. Further calculations addressing more exclusive
observables remain to be confronted with experimental data on a 
quantitative level.

\end{document}